\documentclass[aps,twocolumn,superscriptaddress,showpacs,nofootinbib]{revtex4}

\usepackage{amsfonts,amssymb,amsmath,bm}
\usepackage{graphicx}

\usepackage{bm}

\renewcommand{\vec}[1]{{\bm #1}}
\newcommand{\ul}{{\bm u}_{\scriptscriptstyle \mathrm{L}}}
\newcommand{\deltal}{{\delta}_{\scriptscriptstyle \mathrm{L}}}

\newcommand{\ue}{\mathrm{e}}
\newcommand{\ui}{\mathrm{i}\,}

\begin{document}
\draft{Submitted to \textit{Physica D} (Proceedings of EE250)}

\title{Complex-space singularities of 2D Euler flow in Lagrangian coordinates}

\author{Takeshi Matsumoto}
\email{takeshi@kyoryu.scphys.kyoto-u.ac.jp}
\affiliation{Dep.\  Physics, Kyoto Univ., Kitashirakawa Oiwakecho Sakyoku Kyoto, 606-8502, Japan}
\affiliation{Labor.\ Cassiop\'ee, UNSA, CNRS, OCA, BP~4229, 06304 Nice Cedex
  4, France}

\author{J\'er\'emie Bec}
\affiliation{Labor.\ Cassiop\'ee, UNSA, CNRS, OCA, BP~4229, 06304 Nice Cedex 4, France}

\author{Uriel Frisch}
\affiliation{Labor.\ Cassiop\'ee, UNSA, CNRS, OCA, BP~4229, 06304 Nice Cedex 4, France}

\begin{abstract}
We show that, for two-dimensional space-periodic incompressible flow,
the solution 
can be evaluated numerically  in Lagrangian  coordinates  with the same
accuracy achieved in standard Eulerian spectral methods. This allows the
determination of complex-space Lagrangian singularities. Lagrangian
singularities are found to be closer to the real domain than Eulerian
singularities and seem to correspond to fluid particles which escape
to (complex) infinity by the current time. Various mathematical conjectures
regarding Eulerian/Lagrangian singularities are presented.
\end{abstract}



\maketitle

\section{\label{intro}Introduction}
Solutions to the incompressible Euler equation, starting from
entire initial data (e.g.\ trigonometric polynomials), can
be analytically continued to the complex space
as long as they stay analytic in the real space.  
Furthermore it is known since the seventies that
any singularities in the real space, if they exist,
have to be preceded by complex-space singularities 
\cite{Benachoura,Benachourb}.
In 2D Euler flows, it is known that initial real-space
analyticity for periodic solutions is never lost. This was proven in
Refs.~\cite{hoelder,wolibner} in which it was shown that the distance $\delta(t)$ 
to the real domain 
of the nearest complex-space  singularity, measured by the  exponential
falloff of the Fourier amplitude, 
decreases at large times  at most as a double exponential. 
Actually, already 
twenty five years ago, spectral simulations with $256^2$ Fourier modes
indicated that the  decrease is more like a simple exponential \cite{tracing}. 
Spectral simulations 
at much higher resolutions, up to $8192^2$, which will be published
elsewhere also indicate  a
behavior  much closer to a
single than to a double exponential.\footnote{We nevertheless conjecture that 
by suitable analytic regularization of the flow considered in
Ref.~\cite{bahouri-chemin} long-lasting transients with a double exponential
decrease can be achieved.}
The discrepancy between the simple and the double exponential is generally
believed to be due to the phenomenon of \textit{depletion}: the flow
organizes itself into ribbon-like vortical structures in which the
nonlinearity is almost completely suppressed (the nonlinearity would vanish
identically if the flow depended on a single Cartesian coordinate). The same
phenomenon also exists in three dimensions and could conceivably 
prevent finite-time blow up. 

In three dimensions the Beale--Kato--Majda (BKM) theorem implies that
any blow up must be accompanied by the unboundedness of the modulus of
the vorticity in the real domain \cite{bkm} (see also \cite{mb}).  In
two dimensions, when the initial vorticity is bounded, this is of
course ruled out by vorticity conservation. More precisely, it is
ruled out in the real domain, but not in the complex domain. Actually,
increasingly strong numerical evidence has been obtained indicating
that the vorticity is infinite at complex singularities \cite{blue,
fdr, physd}.\footnote{We have tried but failed to derive such a result
from a complex version of the the BKM argument.} Such numerical results
were obtained only for flows in which the initial stream function is a
trigonometric polynomial (the 2D analogues of the famous Taylor--Green
flow \cite{tg}), which are instances of entire functions, that is,
analytic functions that have no singularity at finite complex
locations.

The ``experimental result'' about infinite vorticity along the complex
singularities in two dimensions has an important consequence: because
the conservation of vorticity along fluid particle trajectories
carries over to complex trajectories, (Eulerian) complex locations
with infinite vorticity are associated with fluid particles initially
at complex infinity; indeed, this is the only place where an entire
function can be infinite.  We were thus led to investigate the issue
of (complex) singularities in Lagrangian coordinates. A Lagrangian
singularity is a location at which the (analytic continuation of the)
Lagrangian map goes singular. Could it be that for two-dimensional
flow there are no (complex) Lagrangian singularities at finite
distance?  In other words: does the flow in Lagrangian coordinates
preserve its initial entire character?   A few years ago we performed very
accurate numerical simulations, reported here for the first time, and
we found strong evidence that the answer is ``no''. W.~Pauls and one
of us (TM) \cite{PM} then found a very simple counterexample to the
preservation of the entire character: the ``AB flow'' $\psi = \sin x_1
\cos x_2$ is an entire steady solution to the 2D Euler equation in
Eulerian coordinates.  For this flow, the trajectories of fluid
particles can be expressed by elliptic functions and it was shown
that, for any real positive time $t$, there exist complex
initial locations of fluid particles which are mapped to infinity
at time $t$  and which thus are Lagrangian singularities.

There is a considerable renewal of interest in the Lagrangian
structure of flows, both from a theoretical and experimental point of
view (such issues frequently came up during the Euler conference).  It
is thus of interest to show that the Lagrangian description of flows
can be obtained numerically with an accuracy comparable to that
available by spectral methods for the Eulerian description. The
present paper is organized as follows. In Section~\ref{s:lagaccurate}
we describe two numerical algorithms, which can be used for Lagrangian
integration. In Section~\ref{s:results} we apply this to the
identification of complex Lagrangian singularities. Here, all 
numerical studies are presented for the (unsteady) 2D flow with the
simple initial condition
\begin{equation}
  \psi_0 = \cos x_1 + \cos 2 x_2\;,
   \label{e:init}
\end{equation}
which has been used in Refs.~\cite{blue, fdr, physd}; key results are
also checked with the flow 
\begin{equation}
  \psi_0 = \cos x_1 + \cos 2 x_2 +\sin(2x_1-x_2)\;,
   \label{e:init2}
\end{equation}
which has less symmetry than \eqref{e:init}.
Some concluding remarks, with emphasis on mathematical conjectures,  
are presented in Section~\ref{s:conclusion}.

\section{Numerical solution in Lagrangian coordinates with spectral 
accuracy}
\label{s:lagaccurate}

Our goal here is to obtain the velocity field ${\bm u}$ as  a function of the
Lagrangian location ${\bm a}$ and the time $t$. This Lagrangian 
field will be denoted $\ul({\bm a},\,t)$.

With  simple boundary conditions, e.g. spatial periodicity, the easiest
way to obtain high accuracy in a Eulerian simulation is to use a spectral
or pseudo-spectral method \cite{go}. For analytic flow, whose
Fourier transform decreases exponentially at high wavenumbers, the truncation 
error will then also decrease exponentially with the resolution.

How does one carry this over to Lagrangian coordinates? In principle one
can write an integro-differential equation for the (time-dependent) 
Lagrangian map ${\bm a} \mapsto {\bm x}$. This equation has however
nonlinearities with  denominators which are not easily handled numerically.

We present here two alternative methods, the spectral particle-tracking method
(Section~\ref{s:tracking}) and the spectral displacement-Newton method
(Section~\ref{s:dn}).

\subsection{Particle tracking  method}
\label{s:tracking}

Obviously, the Lagrangian velocity field can be obtained by composing
the Eulerian velocity field ${\bm u}({\bm x},\,t)$ with the Lagrangian
map $ {\bm x} ({\bm a},t)$. The former can be obtained by standard
spectral integration. The latter is  the solution of the characteristic
equation
\begin{equation}
 \partial_t{\vec{x}}(\vec{a},\, t) = 
\vec{u}(\vec{x}(\vec{a},\,  t),\,\, t), \qquad \vec{x}(\vec{a},\, 0 )=
\vec{a}\;.
\label{e:lagparticle}  
\end{equation}
In the tracking method, we select a uniform grid of Lagrangian points
and ``track'' the fluid particles by integrating \eqref{e:lagparticle} 
along all the relevant fluid particle trajectories. This can be done, e.g.
using a fourth-order Runge-Kutta method. The problem is that, even 
if the initial positions coincide with Eulerian collocation points,
this usually ceases to hold subsequently. 
Hence the Eulerian field must be interpolated.
In order not to loose the spectral accuracy, the interpolation can
be done using the Fourier series representation
\begin{equation}
 \vec{u}(\vec{x},\, t) = \sum_{\vec{k}} \hat{\vec{u}}(\vec{k}, t) 
\,\ue ^{\ui 
\vec{k}\cdot\vec{x}}\;.
\label{e:fourier}
\end{equation}
A difficulty is that, since the relevant ${\bm x}$'s are not
collocation points, the velocities given by \eqref{e:fourier} cannot
be evaluated using fast Fourier transforms but must be calculated
``na\"{\i}vely'' in $O(N ^4)$ operations if we use an $N \times N$
grid. Furthermore this has to be done at every time step.  Since the
number of time steps needed to reach a given time $t$ order unity is
proportional to the resolution $N$, this method has a fairly large
computational complexity $O(N ^5)$ and thus also a significant
accumulation of round-off errors.  For large values of the resolution
$N$ (512 or more) the particle tracking method is not very practical
unless we restrict the Lagrangian grid to being much coarser than the
Eulerian grid.

\subsection{Displacement-Newton method}
\label{s:dn}

This method makes use of the fact that the \textit{inverse Lagrangian map}
${\bm a}({\bm x},\, t)$ satisfies, in Eulerian coordinates, the equation
\begin{equation}
\partial_t \vec{a} + \vec{u(\vec{x},t)}\cdot\nabla \vec{a} =0\;,
 \label{e:aeq}
\end{equation}
which just expresses the constancy of the Lagrangian  location ${\bm a}$ under
advection by the velocity field. This equation can be solved along with
the basic Euler equation, both in Eulerian coordinates. This will however
yield a map which still has to be inverted to obtain the direct
Lagrangian map. 

For periodic boundary conditions the direct and inverse maps are not
periodic and it is more convenient to work with the displacement field,
here defined as
\begin{equation}
\vec{d}(\vec{x}, t) \equiv \vec{a}(\vec{x}, t) - \vec{x}\;.
\label{e:defd}
\end{equation}
It follows from \eqref{e:aeq} and \eqref{e:defd} 
that the displacement satisfies the following
equation in the Eulerian coordinates
\begin{equation}
 \partial_t \vec{d}(\vec{x}, t) + (\vec{u}(\vec{x}, t)\cdot\nabla)\vec{d}(\vec{x}, t) 
= -\vec{u}(\vec{x}, t),
\label{e:deq}  
\end{equation}
with the initial condition $\vec{d}(\vec{x}, 0) = {\bm 0}$. This equation
can be solved along with the Euler equation to obtain the displacement in
Eulerian coordinates on a uniform grid of $N\times N$ collocation points.

Then comes the difficult step, namely the inversion. For this we define
the off-grid displacement, as above,  by its Fourier series, extended off-grid
and we try to find the ${\bm x}$ locations  associated to a set of Lagrangian
collocation points on the regular grid $\vec{A} = (2\pi i/N, 2\pi j /N),\,
i, j = 0,\ldots N - 1$. We then determine the direct Lagrangian map
${\bm x}({\bm A},\, t)$ as the solution of the equation
\begin{equation}
\vec{d}(\vec{x}, t) =  \vec{A}(\vec{x}, t) - \vec{x}\;.
\label{e:target}
\end{equation}
First we determine an approximate on-grid solution ${\bm X}({\bm A},\, t)$ by
  finding from the inverse map the ${\bm a}$ point nearest to ${\bm
  A}$ and its inverse Lagrangian antecedent ${\bm X}$. We then set $\vec{x} = \vec{X} + \delta \vec{x}$ and refine the
  solution
of \eqref{e:target} by using a standard Newton method. This requires the calculation of off-grid
values of derivatives, which are again obtained from ``na\"{\i}ve''
evaluations of the corresponding Fourier series
\begin{equation}
 \frac{\partial \vec{d}}{\partial x_j} = \sum_{\vec{k}} \ui k_j
 \hat{\vec{d}}(\vec{k},\, t) \,\ue ^{\ui \vec{k}\cdot\vec{x}}\;,
\label{e:slowFT}
\end{equation}
where $\hat{\vec{d}}(\vec{k},\, t)$ are the Fourier coefficients of the
displacement (evaluated in Eulerian coordinates).  For each stage of the
Newton iteration $O(N^4)$ operations are required. The number of stages needed
to achieve an accuracy $\epsilon$ consistent with double precision is
typically five. If the number of output times at which we want to evaluate the
Lagrangian velocity field is much smaller than the resolution $N$, the
displacement-Newton method is much faster than particle tracking.

\section{Results}
\label{s:results}

We have applied the two methods described in the previous section to
the flow with the initial condition \eqref{e:init}. The methods give 
consistent results but the highest resolution (here $N=512$) is more
easily achieved  with the displacement-Newton method, which has been used
to obtain the results reported here.

The solution  of the Euler equation 
\begin{equation}
\partial {\bm u} + {\bm u}\cdot
\nabla {\bm u} = -\nabla p, \quad \nabla \cdot {\bm u}=0\;,
\label{euler}
\end{equation} 
together with the displacement equation \eqref{e:deq} was obtained
by a standard pseudo-spectral method with two-thirds dealiasing
and a fourth-order Runge--Kutta temporal integration.

Then we applied the displacement-Newton method (with five iterations)
and  $\epsilon =10^{-14}$. The results were checked by computing 
the Lagrangian vorticity, which 
should be equal to its initial value for 2D Euler flow, and was indeed
found to be so with an accuracy of $10^{-10}$.

In order to locate complex-space singularities for the Lagrangian 
solution, we applied the tracing method \cite{tracing}: the Lagrangian solution is
represented by its Fourier series 
\begin{equation}
 \ul(\vec{a},\, t) = \sum_{\vec{k}} \hat{\ul}(\vec{k}, t) \,\ue ^{\ui\vec{k}\cdot\vec{a}}\;.
\label{e:lagF}
\end{equation}
Then  the following asymptotic representation is used for the shell-summed  
high wavenumber Fourier amplitudes (\cite{fdr})
\begin{equation}
  \sum_{k \le |\vec{k}| < k + 1} |\hat{\ul}(\vec{k}, t)|
   \simeq C(t) k^{\alpha(t)}\exp[-\deltal(t) k]\;.
\end{equation}
Here $\deltal(t)$ is the width of the Lagrangian analyticity strip, that
is the distance at time $t$ from the real domain of the nearest (Lagrangian)
complex-space  singularity. The same analysis is applied also to the
Eulerian velocity.

\begin{figure}
\includegraphics[scale=0.7]{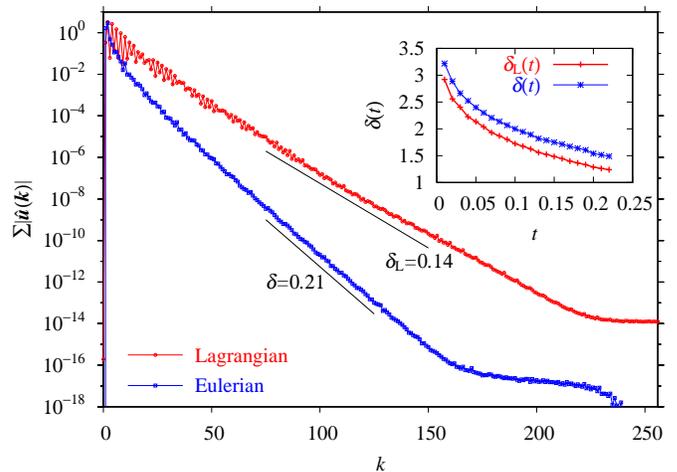}
\caption{Shell-summed amplitudes of Eulerian and Lagrangian velocities at
time $t = 1.245$ in lin-log coordinates. The initial velocity  is given by
\eqref{e:init}.  Inset: time variation (at short times) of the width
of the analyticity strip in Eulerian coordinates ($\delta(t)$) and
Lagrangian  coordinates ($\deltal(t)$).}
\label{f:one}
\end{figure}

Figure~\ref{f:one} shows the wavenumber dependence of the shell-summed
amplitudes for the Eulerian and Lagrangian velocities at the same time, chosen
in such a way that there is only a modest range of wavenumbers at high
$k\equiv|{\bm k}|$ at
which the rounding errors swamp the (roughly) exponential signal.  They both
exhibit exponential decay from which the Eulerian $\delta(t)$ and its
Lagrangian counterpart $\deltal(t)$ are measured.  It is seen that the
Lagrangian $\delta$ is significantly smaller than the Eulerian
one. Actually, $\deltal(t)< \delta(t)$ holds for all times 
$t<1.245$ (the latest time analyzed). We also checked that the inequality
$\deltal< \delta$ holds for the flow with the initial condition 
\eqref{e:init2} which has less symmetry than \eqref{e:init}.

The exponential decay with the wavenumber $k$ of the shell-summed
Lagrangian Fourier amplitude
is strong evidence that there are singularities of the Lagrangian
velocity 
$\ul({\bm a},\, t)$
at a finite distance from the real domain;  thus it cannot be
an entire function.  We also obtained numerical evidence that the
Lagrangian map has the same locations  of complex Lagrangian 
singularities as the Lagrangian velocity and that the inverse  Lagrangian map 
has the same locations  of complex Eulerian
singularities as the Eulerian velocity. For the very simple Eulerian
steady flow
investigated in Ref.~\cite{PM}, Lagrangian singularities
are mapped to Eulerian (complex) infinity. Is this also the case for
the present flow which has non-trivial Eulerian dynamics? Here, the
answer
appears to be ``yes''.
Specifically, let ${\bm a}_\star$ be a Lagrangian singular location
corresponding
to time $t>0$,
say the one closest to the real domain or one near this position.
Does ${\bm d}({\bm a},\,t) \equiv {\bm a} -{\bm x}({\bm a},\,t)$ tend to infinity as ${\bm a}\to {\bm
  a}_\star$? In principle we can find the scaling law of any component
of  ${\bm d}$ as ${\bm a}\to {\bm   a}_\star$, if we have sufficiently
accurate high-resolution data for the Fourier transform of ${\bm
  d}({\bm a})$ at high wavenumbers. This is explained in Section~4.2 of Ref.~\cite{physd}.
It requires the determination not only of the exponential decrement
$\deltal$ but of the exponent of the  algebraic prefactor in front of
the 
exponential which controls the nature of the singularity in complex
${\bm a}$-space. With a resolution
of only $512^2$, such exponents are rather poorly determined. It is
likely
that both components of ${\bm
  d}({\bm a})$ blow up as $s^{-\beta}$ where $s$ is the modulus
of ${\bm a}- {\bm   a}_\star$ and the exponent $\beta$ is about $3/2$
but with and  error bar so large that a negative value cannot be
completely ruled out.\footnote{We have applied the same method of analysis to
  the behavior of ${\bm d}({\bm x})$ when approaching a Eulerian singularity 
at ${\bm x}_\star$. The displacement seems again to diverge with and exponent 
$\beta$ around $3/2$ (implying also the
divergence of the Eulerian vorticity) but the quality of the scaling is
again dubious.} We shall revisit such issues from a theoretical
point of view in the concluding section.

\section{Concluding remarks}
\label{s:conclusion}

\begin{figure}
\includegraphics[scale=0.9]{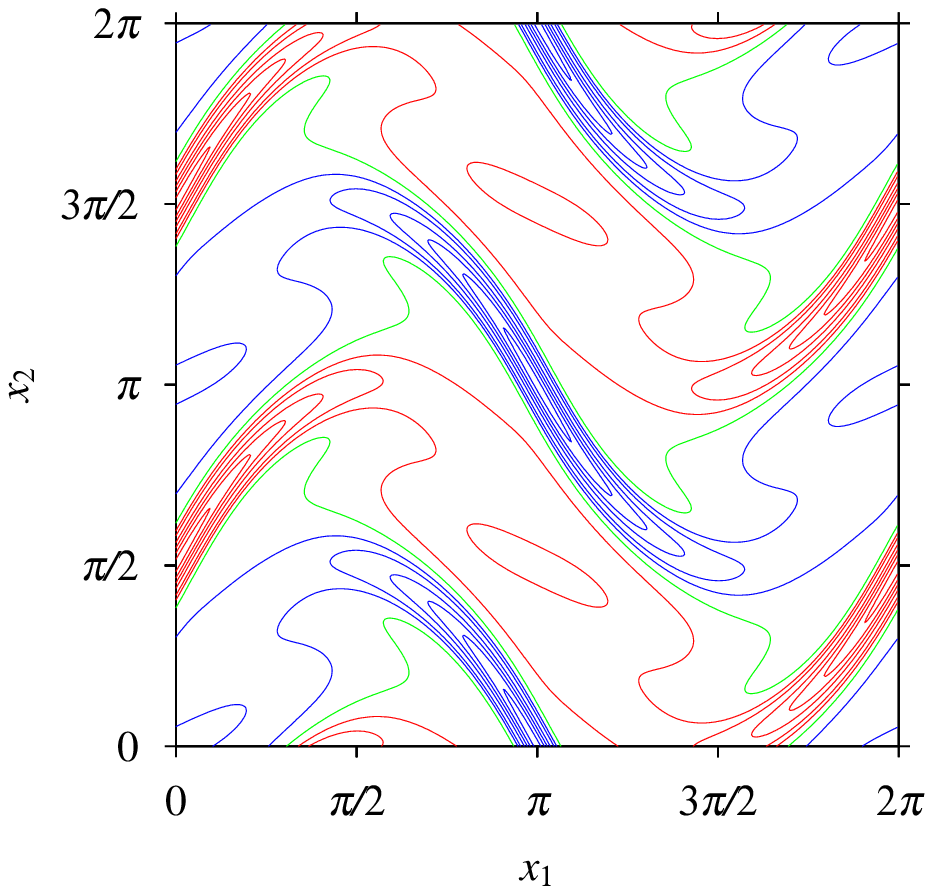}\\[4ex]
\includegraphics[scale=0.9]{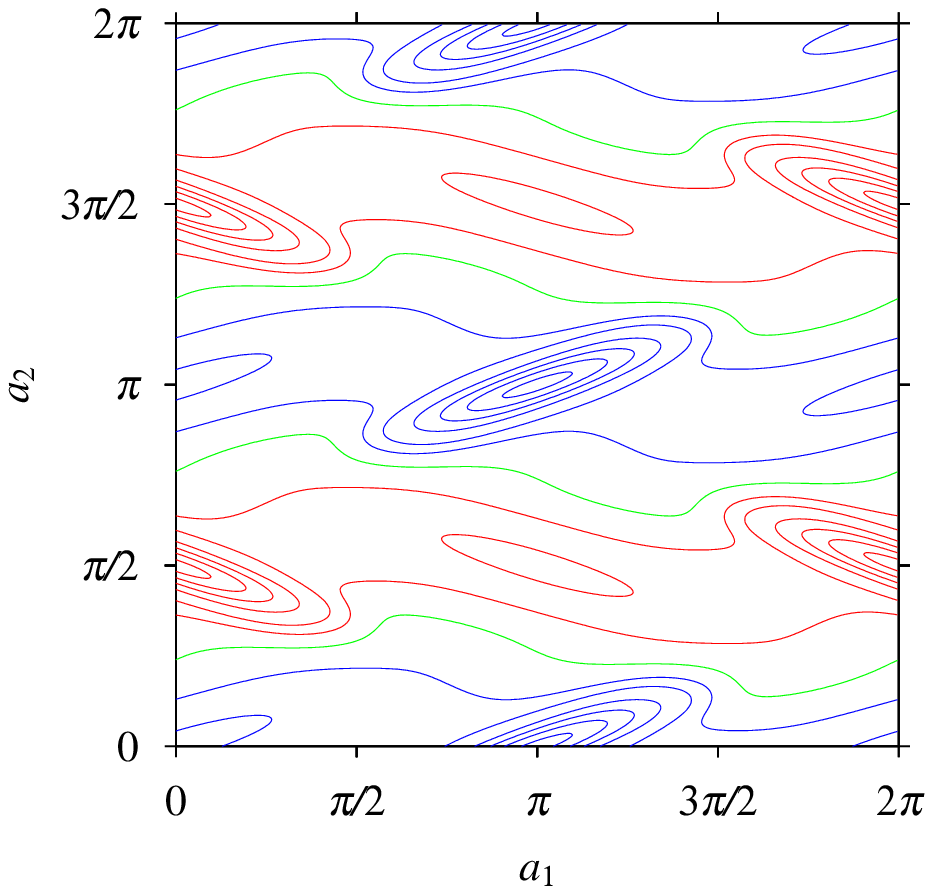}
\caption{Contours of the Laplacian of the vorticity shown for the same flow as
 in Figure~\ref{f:one}, shown at  $t = 1.245$. Upper figure: Eulerian
coordinates; lower figure: Lagrangian coordinates. For both figures  the
 contour values are $0,\,\,\pm 12.5,\,\,\pm 25,\,\, \pm 50,\,\,\pm 75,
\,\,\pm 100,
\,\,\pm 125$; red: positive, green: zero, blue: negative.}
\label{f:two}
\end{figure}

We have shown that the  simple 2D incompressible non-steady flow with
the initial condition \eqref{e:init} has complex singularities not
only in Eulerian but also in Lagrangian coordinates. The Lagrangian
singularities are significantly closer to the real domain than the
Eulerian ones. 
A possible interpretation of this was
given by S.~Orszag (private communication 2003): in Eulerian
coordinates the build up of singularities is  slowed down
by the aforementioned phenomenon of depletion, whereas in Lagrangian
coordinates any flow which is non-uniform will keep changing
non-trivially, even if it is steady in Eulerian coordinates. 
To illustrate this we have shown in Figure~\ref{f:two} the  (Eulerian)
Laplacian of the
vorticity $\nabla ^2 \omega$ in both Eulerian and Lagrangian coordinates.
The former representation displays strongly depleted ribbon-shaped structures, 
not seen in the latter.

Now we wish to comment on the results concerning the analytic
structure in Lagrangian coordinates and on possible generalizations to
other 2D flow with space-periodic entire initial data. The most
obvious result is that, since the vorticity remains unchanged along
fluid particle trajectories in 2D, the Lagrangian vorticity field
stays entire for all times and thus devoid of any singularities other
than at complex infinity. The Lagrangian velocity field and the
Lagrangian map both have complex singularities (presumably along
one-dimensional
complex manifolds)  and the numerical evidence is
that these are at the same locations. Proving this partially can 
perhaps be done by writing the velocity in terms of the
vorticity using the (periodicity-modified) Biot--Savart integral
representation
and then  making the change of variable from Eulerian to Lagrangian
coordinates. On the resulting integral, using the fact that the
initial vorticity is entire, it may be possible to show that if the 
Lagrangian map ${\bm x}({\bm a},\, t)$ is analytic for some (complex)
${\bm a}$, the same holds for the Lagrangian velocity. 

One of the most striking results reported in Section~\ref{s:results},
but one for which the evidence is a bit shaky, is that Lagrangian
singularities at time $t>0$ correspond to fluid particles which at
time $t$ escape to  infinity. Here are some observations which could be
useful in  proving this. The idea is to show that there is a contradiction if
at time $t>0$  a Lagrangian singularity ${\bm a}_\star$ at a finite location
is mapped
to a point ${\bm x}_\star$ which is not at infinity. 
Indeed, if ${\bm x}_\star$ is at finite distance, from the fact that the 
Jacobian of the Lagrangian map
is one, it follows that  ${\bm x}_\star$ must be a singularity of the
inverse Lagrangian map ${\bm x} \mapsto {\bm a}$. The Eulerian
vorticity can be obtained by composing the inverse Lagrangian map and
the initial (entire) vorticity. Composing a function singular at ${\bm
  x}_\star$ with one which is entire does not necessarily yield a singular
function. Perhaps with some extra work it can be proved that the Eulerian
vorticity is indeed singular at ${\bm x}_\star$. We already pointed out
in the Introduction that for 2D space-periodic initially entire flow 
there is  numerical evidence that the vorticity 
is infinite at (complex) Eulerian singularities. If this can also be proved,
it then follows that ${\bm a}_\star$ is at infinity and thus we 
have a contradiction.

The global picture emerging from all this is (tentatively) the
following: for entire periodic initial data in 2D, the solutions
of the incompressible Euler equation have complex Eulerian
singularities corresponding to fluid particles initially at infinity
and Lagrangian singularities corresponding to fluid particles
currently at infinity. In both coordinates singularities correspond to
some particle escaping  to infinity; this mechanism for incompressible 
fluids is very different from the one operating for the one or multi-dimensional
\textit{compressible} Burgers equation for which singularities are mostly
associated to the vanishing of the Jacobian of the the Lagrangian map
(see, e.g., Ref.\cite{bec-khanin}).  

We cannot at present rule out
that the same
scenario holds in three dimensions but it may not be consistent with
real blow up. Of course, there are major differences in 3D; for example,
vorticity is not conserved. However, the Lagrangian numerical techniques
presented in this paper are easily extended to the three-dimensional
case.

\begin{acknowledgments}
We are very grateful to W.~Pauls for many useful
remarks. Thanks are also due to C.~Bardos and S.~Orszag.
TM was supported by
the Grant-in-Aid for the 21st Century COE ``Center for Diversity and
Universality in Physics'' from the Japanese Ministry of Education,
by the Japanese Ministry of Education Grant-in-Aid for Young
Scientists [(B), 15740237, 2003] and by the French Ministry of Education.
\end{acknowledgments}

\end{document}